\documentclass[fontsize=11pt,headings=small,headinclude=true,footinclude=true, 
 headings=optiontohead,toc=listof,DIV=22,listof=leveldown, headsepline , letterpaper]{scrartcl} 
\usepackage[utf8]{inputenc}
\usepackage[T1]{fontenc}
\usepackage[ngerman, english]{babel}

\newcommand{\dptitle}{Helmholtz Decomposition and Rotation Potentials \par in n-dimensional Cartesian Coordinates}
\newcommand{\dptitleclean}{Helmholtz Decomposition and Rotation Potentials in n-dimensional Cartesian Coordinates}

\newcommand{\dpautoren}{Erhard Glötzl$^1$ \href{https://orcid.org/0000-0002-3092-8243}{\includegraphics[width=1em]{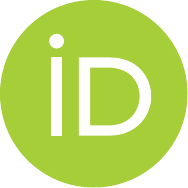}}, Oliver Richters$^{2,3}$ \href{https://orcid.org/0000-0001-8253-4716}{\includegraphics[width=1em]{orcid_logo.pdf}}}
\newcommand{\dpautorenclean}{Erhard Glötzl, Oliver Richters}

\newcommand{\dpaffiliation}{\small 1: Institute of Physical Chemistry, Johannes Kepler University Linz, Austria. \\ 2: ZOE. Institute for Future-Fit Economies, Bonn, Germany. \\ 3: Department of Business Administration, Economics and Law, \\ Carl von Ossietzky University, Oldenburg, Germany.}

\usepackage{amsmath}
\usepackage{amssymb,amsthm,amsfonts,textcomp}
\usepackage{color, transparent}
\usepackage{dpfloat}
\usepackage{array}
\usepackage{txfonts}
\usepackage{booktabs, multirow}
\usepackage{paralist}
\usepackage{hhline}
\usepackage{graphicx}
\usepackage{hanging}
\usepackage{bm} \usepackage{bbold}
\usepackage[figuresleft]{rotating}
\usepackage{acronym}
\usepackage{microtype}
\usepackage[hang]{footmisc}
\usepackage{changepage} 
\usepackage{appendix}
\usepackage{csquotes}
\usepackage{pdfpages}
\usepackage{xcolor}

\usepackage{pifont} 

\usepackage[backend=biber, natbib=true, style=authoryear-comp, sorting=nyt, url=true, isbn=false, doi=true, eprint=true, maxcitenames=2, citetracker=true, uniquename=false, labeldate=year]{biblatex}

\DeclareSourcemap{
 \maps{
  \map{
   \step[fieldsource=language, fieldset=langid, origfieldval, final]
   \step[fieldset=language, null]
  }
 }
}

\AtEveryBibitem{%
   \clearfield{day}%
   \clearfield{month}
   \clearfield{endday}%
   \clearfield{endmonth}%
}

\usepackage[inner=3.2cm,outer=3.2cm,top=1.2cm,bottom=0.8cm,includeheadfoot , heightrounded]{geometry}

\usepackage{chngcntr}

\makeatletter
\newcommand\arraybslash{\let\\\@arraycr}
\makeatother

\makeatletter
\g@addto@macro\UrlBreaks{\do\*\do\~\do\'\do\"\do\a\do\b\do\c\do\d\do\e\do\f\do\g\do\h\do\i\do\j\do\k\do%
\l\do\m\do\n\do\o\do\p\do\q\do\r\do\s\do\t\do\u\do\v\do\w\do\x\do\y\do\z\do\&\do\1\do\2\do\3\do\4\do\5\do\6\do\7\do\8\do\9\do\0\do\.}
\makeatother

\usepackage{scrlayer-scrpage}

\clearscrheadfoot
\ihead{\pagemark}
\ohead{\scriptsize Glötzl, Richters: \dptitleclean}
\let\OLDthebibliography\thebibliography
\renewcommand\thebibliography[1]{
 \OLDthebibliography{#1}
 \setlength{\parskip}{0pt}
 \setlength{\itemsep}{0pt plus 0.3ex}
}

\KOMAoption{listof}{leveldown}

\usepackage{ragged2e}
\newcolumntype{L}[1]{>{\raggedright\arraybackslash}p{#1}} 
\newcolumntype{C}[1]{>{\centering\arraybackslash}p{#1}} 
\newcolumntype{R}[1]{>{\raggedleft\arraybackslash}p{#1}} 

\DeclareMathOperator{\grad}{grad}
\DeclareMathOperator{\Div}{div}
\DeclareMathOperator{\curl}{curl}

\DeclareMathOperator{\ROT}{ROT}
\DeclareMathOperator{\Tr}{Tr}

\newcommand{\newtonintegralt}{\textstyle ^N\!\!\!\!\int} 
\newcommand{\newtonintegral}{\displaystyle ^N\!\!\!\!\!\int}

\newtheorem{theorem}{Theorem}
\newtheorem{lemmarev}[theorem]{Lemma}
\newtheorem{conj}[theorem]{Proposition}
\newtheorem{defin}{Definition}
\newtheorem{corol}[theorem]{Corollary}

\usepackage[hidelinks, breaklinks=true, pdfauthor={\dpautorenclean}, pdftitle={\dptitleclean}, unicode]{hyperref}

\begin{document}

\selectlanguage{english}

\thispagestyle{scrplain}

\begin{center}
 { \Large { \bfseries \sffamily \dptitle \\ \bigskip  \par \par } \vspace{1em} {\large \dpautoren \par \vspace{1em} \normalsize \dpaffiliation \par \vspace{1em} Version 3 -- July 2021 \par } }
\end{center}
\vspace{1.3em}
\begin{addmargin}{0.05\textwidth}

\textbf{Abstract:}
This paper introduces a simplified method to extend the Helmholtz Decomposition to n-dimensional sufficiently smooth and fast decaying vector fields.
The rotation is described by a superposition of $n(n-1)/2$ rotations within the coordinate planes.
The source potential and the rotation potential are obtained by convolving the source and rotation densities with the fundamental solutions of the Laplace equation.
The rotation-free gradient of the source potential and the divergence-free rotation of the rotation potential sum to the original vector field.
The approach relies on partial derivatives and a Newton potential operator and allows for a simple application of this standard method to high-dimensional vector fields, without using concepts from differential geometry and tensor calculus.


\bigskip


\bigskip

\noindent\begin{minipage}[t]{0.84\linewidth}
\textbf{Licence:} Creative-Commons \href{http://creativecommons.org/licenses/by-nc-nd/4.0/}{CC-BY-NC-ND 4.0}. \end{minipage} 
\begin{minipage}[t]{0.15\linewidth}\vspace{-\ht\strutbox}
\includegraphics[width=\columnwidth]{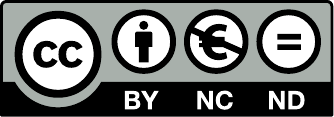}\end{minipage}

\end{addmargin}

\bigskip

\pagestyle{scrheadings}

\section{Introduction}


The Helmholtz Decomposition splits a sufficiently smooth and fast decaying vector field into an irrotational (curl-free) and a solenoidal (divergence-free) vector field.
In $\mathbb{R}^3$, this `Fundamental Theorem of Vector Calculus' allows to calculate a scalar and a vector potential that serve as antiderivatives of the gradient and the curl operator.
This tool is indispensable for many problems in mathematical physics \citep{kustepeli_helmholtz_2016, dassios_uniqueness_2002, sprossig_helmholtz_2009, suda_application_2020}, but has also found applications 
in animation, computer vision, robotics \citep{bhatia_helmholtz-hodge_2013}, or for describing a `quasi-potential' landscapes and Lyapunov functions for high-dimensional non-gradient systems \citep{zhou_quasi-potential_2012, suda_construction_2019}.
The literature review in Section \ref{sec_litreview} summarizes the classical Helmholtz Decomposition in $\mathbb{R}^3$ and the Helmholtz--Hodge Decomposition that generalizes the operator $\curl$ to higher dimensions using concepts of differential geometry and tensor calculus.

Section~\ref{sec_approach} introduces a much simpler generalization to higher dimensions in Cartesian coordinates and states the Helmholtz Decomposition Theorem using novel differential operators.
This approach relies on partial derivatives and convolution integrals and avoids the need for concepts from differential geometry and tensor calculus.
Section~\ref{sec_definitions} defines these operators to derive a source density and a rotational density describing the $\binom{n}{2}$ basic rotations within the 2-dimensional coordinate planes.
By solving convolution integrals, a scalar source potential and $\binom{n}{2}$ rotation potentials can be obtained, and the original vector can be decomposed into a rotation-free vector and a divergence-free vector by superposing the gradient of the source potential with the rotation of the rotation potentials.
Section \ref{sec_properties} states some propositions and proves the theorem. Section \ref{sec_conclusions} concludes.

\section{Literature review}
\label{sec_litreview}
\subsection[Classical Helmholtz Decomposition in R3]{Classical Helmholtz Decomposition in \texorpdfstring{$\mathbb{R}^3$}{R3}}
\label{sec_classic}

In its classical formulation,\footnote{For a historical overview of the contributions by \citet{stokes_dynamical_1849} and \citet{von_helmholtz_uber_1858}, see \citet{kustepeli_helmholtz_2016}.} the Helmholtz Decomposition decomposes a vector field $f \in C^2(\mathbb{R}^3,\mathbb{R}^3)$ that decays faster than $|x|^{-2}$ for $|x| \to \infty$ into an irrotational (curl-free) vector field $g(x) = - \grad \Phi(x)$ with a scalar potential $\Phi \in C^3(\mathbb{R}^3,\mathbb{R})$ and a solenoidal (divergence-free) vector field $r(x) = \curl A(x)$ with a vector potential $A \in C^3(\mathbb{R}^3,\mathbb{R}^3)$ such that $f(x) = g(x) + r(x)$.

The potentials can be derived by calculating the source density $\gamma(x)$ and the rotation density $\rho(x)$:\footnote{Note that if one is not interested in the rotation potentials, $r(x)$ can simply be obtained after determining $G(x)$ and $g(x)$ by calculating $r(x) = f(x)-g(x)$, which was the approach by \citet{stokes_dynamical_1849}.}
\begin{align}
\gamma(x) &= \Div f(x) = \Div g(x), & 
\rho(x) = \curl f(x) = \curl r(x). 
\end{align}
The convolution with the fundamental solutions of the Laplace equation provides the potentials:
\begin{align}
\Phi(x) &= \frac{1}{4\pi} \iiint_{\mathbb{R}^3}  \frac{\gamma(\xi) }{|x-\xi|} d\xi^3, & 
A(x) &= \frac{1}{4\pi}\iiint_{\mathbb{R}^3} \frac{\rho(\xi) }{|x-\xi|} d\xi^3.  \label{eq_potentials_normal}
\end{align}
The Helmholtz decomposition of $f$ is given as: 
\begin{align}
f(x) = g(x) + r(x) \quad \text{ with } \quad  g(x) = - \grad \Phi(x) \quad  \text{ and } \quad  r(x) = \curl A(x).
\end{align}

\subsection{Previous extensions to higher dimensions}

For generalizing the Helmholtz Decomposition to higher-dimensional manifolds, divergence and gradient can straightforwardly be extended to any dimension $n$, but not the operator $\curl$ and the cross product.
This lead to the Hodge Decomposition within the framework of differential forms, defining the operator $\curl$ as the Hodge dual of the anti-symmetrized gradient \citep{vargas_helmholtz-hodge_2014, tran-cong_helmholtzs_1993, mcdavid_generalizing_2006, hauser_fundamental_1970, wolfram_research_curl_2012}.
In two Cartesian dimensions, $\curl$ acting on a scalar field $R$ is a two-dimensional vector field given by
\begin{align}
\curl R(x) = \left[- \frac{\partial R}{\partial x_2}, \frac{\partial R}{\partial x_1} \right] \quad = \left[ \delta_{2k} \frac{\partial R}{\partial x_1} - \delta_{1k} \frac{\partial R}{\partial x_2}; 1 \leq k \leq 2 \right]. \label{eq_2dcurl_scalar}
\end{align}
Square brackets indicate vectors in $\mathbb{R}^n$. The second notation will help to compare this rotation in the $x_1$-$x_2$-plane with the rotation in the $x_i$-$x_j$-plane in Eq.~\eqref{eq_def_ROTij}.
The rotation operator acting on a two-dimensional vector field $f$ yields a scalar field given by:
\begin{align}
\overline{\curl} f(x) = \frac{\partial f_2}{\partial x_1} - \frac{\partial f_1}{\partial x_2} \quad = \frac{\partial f_j}{\partial x_i} - \frac{\partial f_i}{\partial x_j} \text{ with } i = 1, j = 2. \label{eq_2dcurl_vector}
\end{align}
Here, we use the overline to indicate that $\overline{\curl}$ is operating on vector fields, and $\curl$ without overline operates on scalar fields. For $n = 3$, the rotation of a vector field $f$ is usually written as a pseudovector:
\begin{align}
\overline{\curl} f(x) = \left[ \frac{\partial f_3}{\partial x_2} - \frac{\partial f_2}{\partial x_3}, \ \frac{\partial f_1}{\partial x_3} - \frac{\partial f_3}{\partial x_1}, \ \frac{\partial f_2}{\partial x_1} - \frac{\partial f_1}{\partial x_2} \right].
\end{align}
The third component (and analogously first and second) is often understood as the rotation \emph{around} the $x_3$-coordinate.
In order to facilitate the extension to higher dimensions, it should better be discussed as rotation \emph{within} the $x_1$-$x_2$-coordinate plane.
Then, $\curl$ should be understood as an antisymmetric second rank tensor \citep{gonano_cross_2014, mcdavid_generalizing_2006}.
There exist $\binom{n}{2}$ rotations within the coordinate planes.
Only for $n=3$ can each of these rotations be described as a rotation around a vector, as $\binom{n}{2} = n$ if and only if $n = 3$.

For a tensor field $T$ with dimension $n > 3$ and rank $k$, $\curl T$ is a tensor field with dimension $n$ and rank $n - k - 1$.
For a scalar (rank $k = 0$), it consists of $n^{n-1}$ entries.
For a vector field in $\mathbb{R}^n$ (rank $k = 1$), $\overline{\curl}$ consists of $n^{n-2}$ components.
Each component needs $n-2$ indices and is given in Cartesian coordinates by:
\begin{align}
 \left(\overline{\curl} f\right)_{e_1, \dots ,e_{n-2}} &= \sum_{1 \leq l,m \leq n} \tfrac12 \epsilon_{e_1, \dots , e_{n-2}, l, m} \cdot \left( \frac{\partial f_l}{\partial x_m} - \frac{\partial f_m}{\partial x_l} \right) =  \sum_{1 \leq l,m \leq n} -  \epsilon_{e_1, \dots , e_{n-2}, l, m} \cdot \frac{\partial f_l}{\partial x_m}, \label{eq_curl_def}
\end{align}
with the completely antisymmetric Levi--Civita tensor $\epsilon_{\nu_1 \dots \nu_p}$ that is $+1$ (resp. $-1$) if the integers $\nu_1 \dots \nu_n$ are distinct and an even (resp. odd) permutation of $1 \dots n$, and otherwise $0$.

As an example, for $\mathbb{R}^5$, the term $\frac{\partial f_5}{\partial x_4} - \frac{\partial f_4}{\partial x_5}$ can be found with negative sign at positions $_{123}$, $_{231}$ and $_{312}$ and with positive sign at positions $_{132}$, $_{213}$ and $_{321}$. The tensor contains $n(n-1)/2$ different elements apart from sign changes, each repeated $(n-2)!$ times, while the rest of the $n^{n-2}$ elements is zero. This enormous complexity makes higher-dimensional Helmholtz analysis challenging. 

\section{An alternative n-dimensional Helmholtz Decomposition Theorem}
\label{sec_approach}

\begin{figure}[tb]
\hfil\includegraphics[width=0.6\columnwidth]{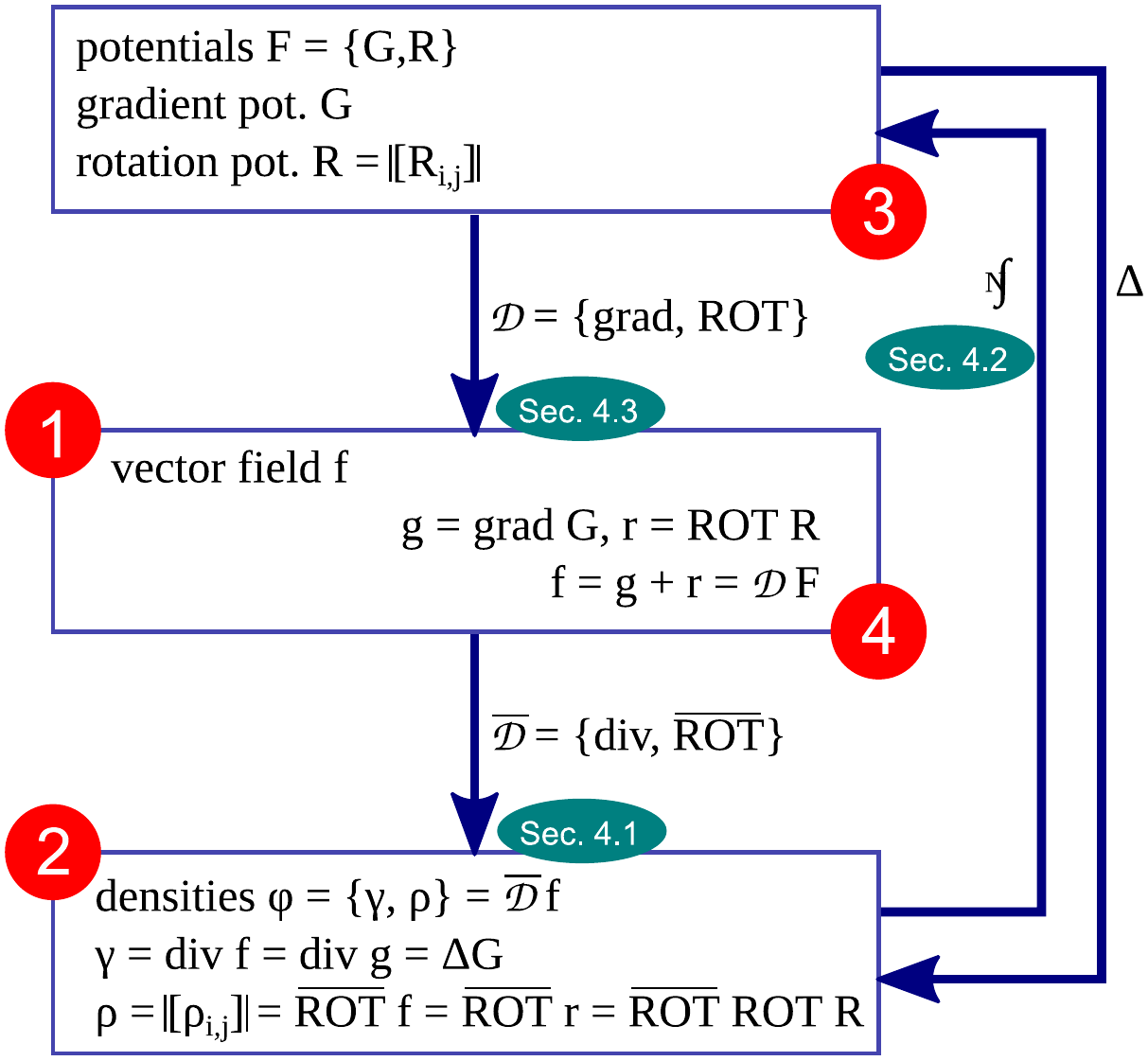}\hfil
\caption{\label{fig_zusammenhang-groessen}Relation between source density $\phi$, vector field $f$ and Newton potential $F$ in $\mathbb{R}^n$. To get the Helmholtz Decomposition of the vector field $f$, derive the densities applying the operator $\overline{\mathcal{D}}$ described in Sec.~\ref{sec_step1}. These densities are convolved with the fundamental solutions of the Laplace equation to derive scalar and rotation potentials (using the `Newton potential operator $\newtonintegralt$') as explained in Sec.~\ref{sec_step2}. The gradient field $g$ and the rotation field $r$ that sum to the original field $f$ are derived using the operator $\mathcal{D}$ defined in Sec.~\ref{sec_step3}.}
\end{figure}

In the following, we present a simpler approach to Helmholtz Decomposition of a twice continuously differentiable vector field $f(x)$ that decays faster than $|x|^{-c}$ for $|x| \to \infty$ and $c > 0$.
The basic idea is to understand the rotation in dimension $n$ as a combination of $\binom{n}{2}$ rotations within the planes spanned by two of the Cartesian coordinates.
In each of these planes of rotation, applying the $\overline{\curl}$ in two dimensions is sufficient.
These $\binom{n}{2}$ rotation densities form the upper triangle of a matrix.
For tractability, we complete this to a $n \times n$ matrix by making it antisymmetric.
It thus contains redundantly both the rotation in the $x_i$-$x_j$-plane and in the $x_j$-$x_i$-plane.
Nevertheless, it has only $n^2$ entries, instead of $n^{n-2}$ entries for $\overline{\curl}$, and contains only the mutually different entries of this operator.


We proceed along the steps shown in Figure~\ref{fig_zusammenhang-groessen}.
Starting {\color{red}\large\ding{182}} from $f$, we calculate {\color{red}\large\ding{183}} the scalar source density $\gamma$ and $n^2$ basic rotation densities $\rho_{ij}$ using the new differential operator $\overline{\mathcal{D}}$, consisting of the well-known divergence $\Div$ and the new operator $\overline{\ROT}$ (Sec.~\ref{sec_step1}).
Each basic rotation density $\rho_{ij}$ corresponds to the rotation within the $x_i$-$x_j$-plane.
The basic rotation densities form the $n^2$-dimensional, antisymmetric rotation density $\rho = \left\llbracket \rho_{ij} \right\rrbracket = \overline{\ROT} f$ and together with $\gamma = \Div f$ the density $\phi = \Big\{ \gamma, \rho(x) \Big\} = \left\{ \gamma, \left\llbracket \rho_{ij} \right\rrbracket \right\}$.
The Newton potential operator $\newtonintegralt$ convolves these densities with the fundamental solutions of the Laplace equation, yielding {\color{red}\large\ding{184}} the scalar `source potential' $G = \newtonintegralt \gamma$ and $n^2$ `basic rotation potentials' $R_{ij} = \newtonintegralt \rho_{ij}$  (Sec.~\ref{sec_step2}).
Similar to the densities, these $n^2$ scalar fields $R_{ij}$ can be jointly written as antisymmetric `rotation potential' $R = \left\llbracket R_{ij} \right\rrbracket$.
With a new differential operator $\mathcal{D}$, combining the well-known gradient $\grad$ with the new operator $\ROT$, operating on the potential $F = \big\{G, R \big\}$, a rotation-free `gradient field' $g = \grad G$ and a source-free `rotation field' $r = \ROT R$ can be calculated (Sec.~\ref{sec_step3}).
In sum {\color{red}\large\ding{185}}, they yield the original vector field $f(x)$.

\section{Definitions}
\label{sec_definitions}

The notation is that square brackets $[f_k]$ indicate a vector, double square brackets $\llbracket R_{ij} \rrbracket$ a $n\times n$ matrix and curly brackets $\left\{ \gamma, \llbracket \rho_{ij} \rrbracket \right\}$ an object of dimension $1+n^2$. For comparison, the table in the Appendix \ref{tabelle-dim-n-3} summarizes the definitions and important properties of the operators and variables, and shows their similarities to the well-known decomposition in dimension 3.

Let $f \in C^2(\mathbb{R}^n,\mathbb{R}^n)$ be a twice continuously differentiable vector field that decays faster than $\left|x\right|^{-c}$ for $\left|x\right| \to \infty$ and $c > 0$.
\begin{align}
f(x) = \left[f_k(x); 1 \leq k \leq n \right] = \left[ f_1(x), \dots, f_n(x) \right].
\end{align}

\begin{defin}[Helmholtz Decomposition, gradient field and rotation field] \label{def_helmholtz}
For a vector field $f \in C^2(\mathbb{R}^n, \mathbb{R}^n)$, a `gradient field' $g \in C^2(\mathbb{R}^n, \mathbb{R}^n)$ and a `rotation field' $r \in C^2(\mathbb{R}^n, \mathbb{R}^n)$ are called a `Helmholtz Decomposition of $f$', if they sum to $f$, if $g$ is gradient of some `gradient potential' $G$, and if $r$ is divergence-free:
\begin{align}
f(x) &= g(x) + r(x), \\
g(x) &= \grad G(x), \\
0 &= \Div r(x).
\end{align}
\end{defin} 

\subsection{From vector field to densities}
\label{sec_step1}

To get from the vector field $f$ to the source and rotation densities, we start with the Jacobian Matrix $J$ of $f$ and its decomposition into a symmetric part $S$ and an antisymmetric part $A$:
\begin{align}
 J = \left\llbracket J_{ij} \right\rrbracket = \left\llbracket \frac{\partial f_i}{\partial x_j} \right\rrbracket = S + A \quad \text{ with } \quad  S = \frac{J + J^\top}{2} \quad \text{ and } \quad A = \frac{J - J^\top}{2}. 
\end{align}

\begin{defin}[scalar source density $\gamma$]
Analogously to $\mathbb{R}^3$, we define the `scalar source density' $\gamma \in C^1(\mathbb{R}^n, \mathbb{R})$ for the vector field $f$ as trace of the Jacobian of $f$:
\begin{align}
\gamma(x) \coloneqq \Tr J = \sum_{i=1}^n \frac{\partial f_i}{\partial x_i} = \Div f(x).
\end{align}
\end{defin}

\begin{defin}[basic rotation density operator $\overline{\ROT_{ij}}$]
We define the `basic rotation density operator' $\overline{\ROT_{ij}}\colon C^1(\mathbb{R}^n,\mathbb{R}^n) \to C^0(\mathbb{R}^n,\mathbb{R})$ for $1 \leq i, j \leq n$ of the vector field $f$ as $2$ times the i-j-element of the antisymmetric part $A$ of the Jacobian $J$:
\begin{align}
 \overline{\ROT_{ij}} f(x) &= 2 A_{ij} =\frac{\partial f_i}{\partial x_j} - \frac{ \partial f_j}{\partial x_i}.
\end{align}
\end{defin}

This generalizes the two-dimensional $\overline{\curl}$ in the $x_1$-$x_2$-plane of Eq.~\eqref{eq_2dcurl_vector} given by $\overline{\curl} f = \frac{\partial f_2}{\partial x_1} - \frac{\partial f_1}{\partial x_2}$ to the rotation in the $x_i$-$x_j$-plane, just with a different sign convention.
To avoid confusion with $\overline{\curl}$, we use $\overline{\ROT}$ for rotation of vector fields -- and later define $\ROT_{ij}$ analogously to $\curl$ without the overline operating on scalar fields.

\begin{defin}[rotation density operator $\overline{\ROT}$] \label{def_overlineROT}
The `rotation density operator' $\overline{\ROT}\colon C^1(\mathbb{R}^n,\mathbb{R}^n) \to C^0(\mathbb{R}^n,\mathbb{R}^{n^2})$ is defined as an antisymmetric operator containing all the basic rotation density operators:
\begin{align}
\overline{\ROT} f(x) \coloneqq \left\llbracket \overline{\ROT_{ij}} f(x) \right\rrbracket = \left\llbracket \left( \frac{\partial f_i}{\partial x_j} -  \frac{\partial f_j}{\partial x_i}\right); 1 \leq i, j \leq n \right\rrbracket.    \label{eq_def_overlineROT}
\end{align}
\end{defin}

\begin{defin}[basic rotational densities $\rho_{ij}$]
We define the $n^2$ `basic rotational densities' $\rho_{ij} \in C^1(\mathbb{R}^n, \mathbb{R}^n)$ for $1 \leq i, j \leq n$ and the matrix $\rho \in C^1(\mathbb{R}^n, \mathbb{R}^{n^2})$ containing these $n^2$~densities as the rotation density operators applied to the vector field $f$:
\begin{align}
\rho_{ij}(x) \coloneqq&\, \overline{\ROT_{ij}} f(x), \\
\rho(x) \coloneqq& \left\llbracket \rho_{ij}(x) \right\rrbracket = \left\llbracket \rho_{ij}(x); 1 \leq i, j \leq n \right\rrbracket = \left\llbracket \overline{\ROT_{ij}} f(x); 1 \leq i, j \leq n \right\rrbracket = \overline{\ROT} f(x). \label{eq_def_rho}
\end{align}
\end{defin}
As $\overline{\ROT_{ij}}$ and $\rho_{ij}$ are antisymmetric, they in fact only contain $\binom{n}{2}$ independent elements, one for each of the $\binom{n}{2}$ coordinate planes. 

\begin{defin}[density derivative $\overline{\mathcal{D}}$, density $\phi$]
We define the `density derivative' $\overline{\mathcal{D}}\colon C^1(\mathbb{R}^n,\mathbb{R}^n) \to C^0(\mathbb{R}^n,\mathbb{R}^{1+n^2})$ of the vector field $f$ that combines $\Div$ and $\ROT$ into one operator, and define the `density' $\phi \in C^1(\mathbb{R}^n, \mathbb{R}^{1+n^2})$ that combines the source and rotation densities:
\begin{align}
\phi(x) \coloneqq \Big\{ \gamma(x), \rho(x) \Big\} = \left\{\gamma(x), \left\llbracket \rho_{ij} (x) \right\rrbracket \right\} = \overline{\mathcal{D}} f(x) \coloneqq \left\{ \Div f(x), \overline{\ROT} f(x) \right\}. \label{eq_def_overlineD}
\end{align}
\end{defin}

\subsection{From densities to potentials}
\label{sec_step2}

The next step derives the potentials starting from the densities using the Newton potential operator.

\begin{defin}[Newton potential operator $\newtonintegralt$]
We define the `Newton potential operator' $\newtonintegralt\colon C^0(\mathbb{R}^n,\mathbb{R}) \to C^2(\mathbb{R}^n,\mathbb{R})$ as convolution of any density $Q \in \{ \phi, \gamma, \rho, \rho_{ij} \}$ that decays faster than $\left|x\right|^{-c}$ for $\left|x\right| \to \infty$ and $c>0$ with the fundamental solutions of the Laplace equation $\Delta K(x) = 0$:
\begin{align}
\newtonintegral Q(x) = \int_{\mathbb{R}^n} K(x, \xi)\ Q(\xi)\ d\xi^n
\end{align}
with the kernel $K(x,\xi)$, using $V_n = \pi^\frac{n}{2} / \Gamma\big(\tfrac{n}{2}+1\big)$ the volume of a unit $n$-ball and $\Gamma(x)$ the gamma function,
\begin{align}
K(x,\xi) &= \begin{cases}
         \frac{1}{2\pi} \left( \log{ | x-\xi | } - \log{ |\xi| } \right) &  n=2,  \\
         \frac{1}{n(2-n)V_n} \left( | x-\xi | ^{2-n} - | \xi | ^{2-n} \right) &  \text{otherwise},
      \end{cases}  \nonumber 
\end{align}
\end{defin}
We call a potential derived this way `Newton potential'.
The decay of $Q$ at infinity guarantees the existence of the convolution integrals, similar to the 3-dimensional case, but the use of the more complex kernel compared to Eq.~\eqref{eq_potentials_normal} allows to relax the restriction to fields decaying faster than $|x|^{-c}$ with $c > 0$, instead of $c=2$ \citep{blumenthal_uber_1905, petrascheck_helmholtz_2015}.\footnote{\citet{tran-cong_helmholtzs_1993} shows that $f(x)$ needs to be bounded at infinity only by $O(\left| x \right| ^l)$ with $l > 0$ a constant using a more complicated convolution integral.
\citet{glotzl_analytical_2021} provide analytical solutions for gradient and rotation potentials and fields of several unbounded vector fields whose components are sums and products of polynomials, exponential and periodic functions.
}

\begin{defin}[source potential $G$]
We define the one-dimensional `source potential' $G \in C^3(\mathbb{R}^n, \mathbb{R})$ as application of the Newton potential operator to the source density:
\begin{align}
G(x) &\coloneqq \newtonintegral \gamma(x) \ \coloneqq \int_{\mathbb{R}^n} K(x, \xi)\ \gamma(\xi)\ d\xi^n, \label{eq_newton_G}
\end{align}
\end{defin}
Note that compared to the case in $\mathbb{R}^3$, we use a different sign convention: $G(x) = -\Phi(x)$.

\begin{defin}[basic rotation potentials $R_{ij}$]
We define $n^2$ `basic rotation potentials' $R_{ij} \in C^3(\mathbb{R}^n, \mathbb{R}) \text{ for } 1 \leq i, j \leq n$, corresponding to the coordinate plane spanned by $x_i$ and $x_j$, as application of the Newton potential operator to the basic rotation densities:
\begin{align}
R_{ij}(x) &\coloneqq \newtonintegral \rho_{ij}(x) = \int_{\mathbb{R}^n} K(x, \xi)\ \rho_{ij}(\xi)\ d\xi^n. \label{eq_newton_Rij}
\end{align}
\end{defin}

\begin{defin}[rotation potential $R$]
We define the `rotation potential' $R \in C^3(\mathbb{R}^n, \mathbb{R}^{n^2})$ as matrix containing the $n^2$ basic rotation potentials $R_{ij}$:
\begin{align}
R(x) &\coloneqq \left\llbracket R_{ij} (x) \right\rrbracket = \left\llbracket R_{ij}(x); 1 \leq i, j \leq n \right\rrbracket = \left\llbracket \newtonintegral \rho_{ij}(x); 1 \leq i, j \leq n \right\rrbracket = \newtonintegral \rho(x). \label{eq_newton_R}
\end{align}
\end{defin}
Here, the Newton potential operator $\newtonintegralt$ is applied in each component.
The rotation potential $R(x)$ is antisymmetric, thus $R_{ij}(x) = - R_{ji}(x)$ and $R_{ii}(x) = 0$, and the rotation potentials therefore contains $\binom{n}{2}$ distinct components.

\begin{defin}[potential $F$]
The `potential' $F \in C^3(\mathbb{R}^{n}, \mathbb{R}^{1+n^2})$ combines source and rotation potentials into one object of dimension $1+n^2$:
\begin{align}
F(x) &\coloneqq \Big\{ G(x), R(x) \Big\} = \newtonintegral \phi(x)  = \newtonintegral \overline{\mathcal{D}} f(x). \label{eq_def_F}
\end{align}
\end{defin}

We know from the theory of the Poisson equation \citep{gilbarg_elliptic_1977} that $\Delta \newtonintegralt Q(x) = Q(x)$ for $Q \in C(\mathbb{R}^n,\mathbb{R}^m)$, which implies:
\begin{align}
\Delta G(x) = \gamma(x), \qquad \Delta R_{ij}(x) = \rho_{ij}(x) \ \forall \ i,j, \qquad \Delta R(x) = \rho(x), \qquad \Delta F(x) = \phi(x). \label{eq_laplace_identities_newton}
\end{align}

\subsection{From potentials to vector fields}
\label{sec_step3}

\begin{defin}[basic rotation operator $\ROT_{ij}$]
We define the `basic rotation operator' $\ROT_{ij}\colon C^1(\mathbb{R}^n,\mathbb{R}) \to C^0(\mathbb{R}^n,\mathbb{R}^n)$ operating on the basic rotation potential $R_{ij}$ with $1 \leq i, j \leq n$ as
\begin{align}
\ROT_{ij} R_{ij}(x) \coloneqq& \left[0, \dots, 0, + \frac{\partial R_{ij}}{\partial x_j}, 0, \dots, 0, - \frac{\partial R_{ij}}{\partial x_i}, 0, \dots, 0 \right]  \label{eq_def_ROTij} \\ \nonumber =& \left[ \delta_{ik} \frac{\partial R_{ij}}{\partial x_j} -  \delta_{jk} \frac{\partial R_{ij}}{\partial x_i}; 1 \leq k \leq n \right],
\end{align}
with the Kronecker delta $\delta_{ik} = 1$ if $i = k$ and $0$ otherwise.
\end{defin}
This operator is a generalization of $\curl$ operating on a scalar field in the two-dimensional case of rotations within the $x_1$-$x_2$-plane in Eq.~\eqref{eq_2dcurl_scalar} given by $\left[ \delta_{2k} \frac{\partial R}{\partial x_1} - \delta_{1k} \frac{\partial R}{\partial x_2}; 1 \leq k \leq 2 \right]$, again with a different sign convention.
It operates in the $x_i$-$x_j$-plane, therefore the non-zero terms in this $n$-dimensional vector are located at positions $i$ and $j$, instead of $1$ and $2$ in the $2$-dimensional case.
Note that $\ROT_{ij} R_{ij}(x) = \ROT_{ji} R_{ji}(x)$, yielding a symmetric object. 

\begin{defin}[rotation operator $\ROT$]
We define the `rotation operator' $\ROT\colon C^1(\mathbb{R}^n,\mathbb{R}^{n^2}) \to C^0(\mathbb{R}^n,\mathbb{R}^n)$ for each rotation potential $R(x) = \left\llbracket R_{ij} (x) \right\rrbracket$ as superposition of the $\binom{n}{2}$ basic rotations in the coordinate planes.
As the antisymmetric potential $R(x)$ of dimension $n^2$ contains each of the $\binom{n}{2}$ basic rotations twice, we have to devide the result by 2.
\begin{align}
\ROT R(x) \coloneqq& \frac12 \sum_{i,j=1}^n \ROT_{ij} R_{ij} (x) = \left[ \sum_{m=1}^n \frac{\partial R_{km}}{\partial x_m}; 1 \leq k \leq n \right]. \label{eq_def_ROT}
\end{align}
\end{defin}
Note that this is identical to $\Div R_{km}$ for an antisymmetric second-rank tensor.

\begin{defin}[gradient field $g$ and rotation field $r$]
The `gradient field' $g \in C^2(\mathbb{R}^n, \mathbb{R}^n)$ and the `rotation field' $r \in C^2(\mathbb{R}^n, \mathbb{R}^n)$ are defined as:
\begin{align}
g(x) &= \grad G(x), \qquad r(x) = \ROT R(x). \label{eq_def_gr}
\end{align}
\end{defin}

\begin{defin}[derivative of the potential $\mathcal{D}$]
We define the `derivative of the potential' $\mathcal{D}\colon C^1(\mathbb{R}^n,\mathbb{R}^{1+n^2}) \to C^0(\mathbb{R}^n,\mathbb{R}^n)$ operating on a potential $F(x) = \{ G(x), R(x) \}$ as the sum of the gradient operating on the source potential $G(x)$ and the rotation operator $\ROT$ operating on the rotation potential $R(x)$:
\begin{align}
\mathcal{D} F(x) &= \mathcal{D} \Big\{G(x), R(x)\Big\} \coloneqq \grad G(x) + \ROT R(x) = g(x) + r(x). \label{eq_def_D}
\end{align}
\end{defin}
We call a potential $F = \{G, R\}$ `antiderivative' of $f$ if $f(x) = \mathcal{D} F(x)$.

Given the conditions on $f(x)$, the potential is uniquely\footnote{\label{footnote_harmonic}Note that by Liouville's theorem, if $H$ is a harmonic function defined on all of $\mathbb{R}^n$ which is bounded above or bounded below, then $H$ is constant, and therefore identical zero if it vanishes at infinity \citep[p.~108]{medkova_laplace_2018}.
Therefore, we do not need to care about integration constants and adding harmonic functions that solve the Laplace Equation $\Delta H(x) = 0$.
If the fields do not decay sufficiently fast, alternative methods to derive Newton Potentials can be found in the literature, see Sec.~\ref{sec_classic}.
They require a careful attention to boundary conditions because $G(x) = \newtonintegralt \Delta (G(x) + H(x))$ with any harmonic function $H(x)$.} determined by Eq.~\eqref{eq_newton_G}.
We will prove that Eqs.~\eqref{eq_def_gr} and \eqref{eq_def_D} are a Helmholtz Decomposition according to Definition~\ref{def_helmholtz}.

\section{Helmholtz Decomposition Theorem}
\label{sec_properties}

\begin{theorem}[Helmholtz Decomposition Theorem] \label{theorem_hd} 

Any twice continuously differentiable vector field $f \in C^2(\mathbb{R}^n,\mathbb{R}^n)$ that decays faster than $|x|^{-c}$ for $\left|x\right| \to \infty$ and $c > 0$ can be decomposed into two vector fields, one rotation-free and one divergence-free.
With the definitions of the operators in Section~\ref{sec_definitions}, let
\begin{align}
G &\in C^3(\mathbb{R}^n,\mathbb{R}), & \text{ \ \ with } G(x) &\coloneqq \newtonintegralt \Div f(x), \\
g &\in C^2(\mathbb{R}^n,\mathbb{R}^n), & \text{ with }\ g(x) &\coloneqq \grad G(x), \\
R &\in C^3(\mathbb{R}^n,\mathbb{R}^{n^2}), & \text{ with } R(x) &\coloneqq \newtonintegralt \overline{\ROT} f(x), \\
r &\in C^2(\mathbb{R}^n,\mathbb{R}^n), & \text{ with }\ r(x) &\coloneqq \ROT R(x),
\end{align}
then
\begin{align}
& \text{$g$, the gradient of $G$, is rotation-free: } & 0 &= \overline{\ROT} g(x) = \overline{\ROT} \grad G(x), \label{eq_hdt1} \\
& \text{$r$, the rotation of $R$, is divergence-free: } & 0 &= \Div r(x) = \Div \ROT R(x), \label{eq_hdt2}  \\
& \text{the potential $\{G, R\}$ is an antiderivative of $f$: } & f(x) &= \mathcal{D}\big\{ G(x), R(x) \big\}, \label{eq_hdt3} \\
& \text{the Helmholtz Decomposition of $f$ is given by: } & f(x) &= g(x) + r(x). \label{eq_hdt4}
\end{align}
\end{theorem}
\begin{proof}
Propositions \ref{conj_gradnorot} and \ref{conj_rotnodiv} show that $g(x)$ is curl-free and $r(x)$ is divergence-free (Eqs.~\ref{eq_hdt1}--\ref{eq_hdt2}).
We introduce and prove Lemma \ref{sec_laplace_identites}, Corollary \ref{sec_corol} and three operator identities as Propositions \ref{conj_laplace_identity1}--\ref{eq_laplace_identity3}.
In Proposition \ref{conj_HT3}, we prove the conditions in Eqs.~(\ref{eq_hdt3}--\ref{eq_hdt4}) that $g(x) + r(x) = \mathcal{D} F(x)$ equals the original vector field $f(x)$.
\end{proof}

\begin{conj}[Eq.~\eqref{eq_hdt1} of the Helmholtz Decomposition Theorem: $g(x) = \grad G(x)$ is rotation-free] \label{conj_gradnorot}
For any twice continuously differentiable vector field $f \in C^2(\mathbb{R}^n, \mathbb{R}^n)$ that decays faster than $\left|x\right|^{-c}$ for $\left|x\right| \to \infty$ and $c>0$, the source potential $G(x) = \newtonintegralt \Div f(x)$ as defined by Eq.~\eqref{eq_newton_G} and its gradient $g(x) = \grad G(x)$ satisfy the following identities:
\begin{align}
\overline{\ROT} g(x) &= \overline{\ROT} \grad G(x) = 0, \\
\overline{\curl} g(x) &= \overline{\curl} \grad G(x) = 0.
\end{align}
\end{conj}
\begin{proof}
Using the Definition \ref{def_overlineROT} of $\overline{\ROT}$ in Eq.~\eqref{eq_def_overlineROT} and of the components of $\overline{\curl}$ in Eq.~\eqref{eq_curl_def}, and permutability of second derivatives:
\begin{align}
\begin{split}  \overline{\ROT} g(x) &= \overline{\ROT} \grad G(x) = \overline{\ROT} \left[ \frac{\partial G}{\partial x_1}, \dots, \frac{\partial G}{\partial x_n} \right] \\ &= \left\llbracket \frac{\partial }{\partial x_i} \frac{\partial G}{\partial x_j} - \frac{\partial }{\partial x_j} \frac{\partial G}{\partial x_i}; 1 \leq i, j \leq n \right\rrbracket = 0, \end{split} \label{eq_grad_no_rot} \\
\begin{split} \left(\overline{\curl} g(x) \right)_{e_1, \dots ,e_{n-2}} &= \sum_{1 \leq i,j \leq n} \tfrac12 \epsilon_{e_1, \dots , e_{n-2}, i, j} \cdot \left( \frac{\partial g_i}{\partial x_j} - \frac{\partial g_j}{\partial x_i} \right) \\
&= \sum_{1 \leq i,j \leq n} \tfrac12 \epsilon_{e_1, \dots , e_{n-2}, i, j} \cdot \left( \frac{\partial}{\partial x_j} \frac{\partial G}{\partial x_i} - \frac{\partial}{\partial x_i} \frac{\partial G}{\partial x_j} \right) = 0. \end{split}
\end{align}
\end{proof}

\begin{conj}[Eq.~\eqref{eq_hdt2} of the Helmholtz Decomposition Theorem: $r(x) = \ROT R(x)$ is divergence-free] \label{conj_rotnodiv}
For any twice continuously differentiable vector field $f \in C^2(\mathbb{R}^n, \mathbb{R}^n)$ that decays faster than $\left|x\right|^{-c}$ for $\left|x\right| \to \infty$ and $c>0$, the rotation potential $R(x) = \newtonintegralt \overline{\ROT} f(x)$ as defined by Eqs.~\eqref{eq_newton_Rij} and \eqref{eq_newton_R} and its rotation $r(x) = \ROT R(x)$ satisfy the following identity:
\begin{align}
\Div r(x) = \Div \ROT R(x) = 0
\end{align}
\end{conj}
\begin{proof}
Using definition of $\ROT$ in Eq.~\eqref{eq_def_ROT} and $\ROT_{ij}$ in Eq.~\eqref{eq_def_ROTij}, exchangeability of derivatives and sums, and permutability of second derivatives:
\begin{align}
\begin{split} \Div \ROT_{ij} R_{ij} (x) =& \Div \left[ \delta_{ik} \frac{\partial R_{ij}}{\partial x_j} - \delta_{jk} \frac{\partial R_{ij}}{\partial x_i}; 1 \leq k \leq n \right] \\
 =& \sum_{k=1}^n \frac{\partial}{\partial x_k} \left(  \delta_{ik} \frac{\partial R_{ij}}{\partial x_j} - \delta_{jk} \frac{\partial R_{ij}}{\partial x_i} \right) 
 =  \sum_{k=1}^n \left( \frac{\partial^2 R_{ij}}{\partial x_i \partial x_j} - \frac{\partial^2 R_{ij} }{\partial x_j \partial x_i} \right) = 0. \end{split} \\
\begin{split} \Div r(x) =& \Div \ROT R(x) = \Div  \sum_{i,j}^n \tfrac12  \ROT_{ij} R_{ij} (x) \\ =& \tfrac12  \sum_{i,j}^n \Div \ROT_{ij} R_{ij} (x) = \sum_{i,j}^n 0 = 0. \label{eq_rot_no_div} \end{split}
\end{align}
\end{proof}

\begin{lemmarev}[Exchangeablity of the Newton potential operator with any partial derivative] \label{sec_laplace_identites}
For any continuously differentiable vector field $Q \in C^1(\mathbb{R}^n, \mathbb{R}^m)$ that decays faster than $\left|x\right|^{-c}$ for $\left|x\right| \to \infty$ and $c>0$, the Newton potential operator as defined by Eq.~\eqref{eq_newton_G} can be exchanged with any partial derivative:
\begin{align}
\dfrac{\partial}{\partial x_k} \newtonintegral Q(x) &= \newtonintegral \frac{\partial Q(x)}{\partial x_k}. \label{eq_HI_partial}
\end{align}
\end{lemmarev}
In the context of this paper, $Q$ can be $G$, $R$, $R_{ij}$, $F$, $f$, $g$, $r$, $\gamma$, $\rho$, $\rho_{ij}$, or $\phi$.
\begin{proof}
\begin{align}
\begin{split} \frac{\partial}{\partial x_k} \newtonintegral Q (x) &= \frac{\partial}{\partial x_k} \int_{\mathbb{R}^n} K(x, \xi) Q(\xi) \,d\xi^n
= \int_{\mathbb{R}^n} Q(\xi) \left( \frac{\partial}{\partial x_k} K(x, \xi) \right) \,d\xi^n \\
&= \int_{\mathbb{R}^n} Q(\xi) \left( - \frac{\partial}{\partial \xi_k} K(x, \xi) \right) \,d\xi^n \end{split} \\
\intertext{using integration by parts in the $k$-component}
&= - \int_{\mathbb{R}^{n-1}} \left( Q(\xi) K(x, \xi) \Big|_{\xi_k = -\infty}^{\infty}  - \int_{\xi_k=-\infty}^{\infty} \frac{\partial Q (\xi) }{\partial \xi_k} K(x, \xi)  \,d\xi_k \right) \,d[\xi_i; i \neq k]  \\
\intertext{(for $n=1$, the outer integral over $\mathbb{R}^0$ in the intermediate step above is omitted) and as in the first summand $Q(\xi) K(x, \xi) \to 0$ for $\xi_k \to \pm \infty$}
&= \int_{\mathbb{R}^n} \frac{\partial Q (\xi) }{\partial \xi_k} K(x, \xi) \,d\xi^n = \newtonintegral \left( \frac{\partial Q(x)}{\partial x_k} \right)\ \forall \ k.
\end{align}
\end{proof}
\begin{corol} \label{sec_corol} Applying Lemma~\ref{sec_laplace_identites} in each component, it follows
\begin{align}
\newtonintegralt \Delta Q(x) &= \Delta \newtonintegralt Q(x) = Q(x) & & \textup{ for } Q \in C^2(\mathbb{R}^n, \mathbb{R}^m),\\
\mathcal{D} \,\newtonintegralt F(x) &= \newtonintegralt \mathcal{D} F(x) & & \textup{ for a potential } F \in C^3(\mathbb{R}^{n}, \mathbb{R}^{1+n^2}), \\
\textup{and } \quad \overline{\mathcal{D}} \,\newtonintegralt f(x) &= \newtonintegralt \overline{\mathcal{D}} f(x) & & \textup{ for a vector field } f \in C^2(\mathbb{R}^n, \mathbb{R}^n).
\end{align}
\end{corol}

\begin{conj}[Operator Identity: $\mathcal{D} \overline{\mathcal{D}} f = \Delta f$] \label{conj_laplace_identity1} 
For any twice continuously differentiable vector field $f \in C^2(\mathbb{R}^n, \mathbb{R}^n)$, the following identity holds:
\begin{align}
\mathcal{D} \overline{\mathcal{D}} f(x) = \grad \Div f(x) + \ROT \overline{\ROT} f(x) = \Delta f(x).
\end{align}
\end{conj}
\begin{proof} The first part follows from the definition of $\mathcal{D}$ and $\overline{\mathcal{D}}$, and Proposition~\ref{conj_gradnorot} and \ref{conj_rotnodiv}. By using Eq.~\eqref{eq_def_ROT} stating $\ROT R(x) = \left[ \sum_m \frac{\partial R_{km}}{\partial x_m}; 1 \leq k \leq n \right]$, it follows:
\begin{align}
\begin{split}
\mathcal{D} \overline{\mathcal{D}} f(x) &= \grad \Div f + \ROT \left\llbracket \left( \frac{\partial f_i}{\partial x_j}  -  \frac{\partial f_j}{\partial x_i}\right); 1 \leq i, j \leq n \right\rrbracket \\
 &= \left[ \frac{\partial}{\partial x_k} \sum_{m=1}^n \frac{\partial f_m}{\partial x_m} + \sum_{m=1}^n \frac{\partial}{\partial x_m} \left( \frac{\partial f_k}{\partial x_m} - \frac{\partial f_m}{\partial x_k} \right) ; 1 \leq k \leq n \right] \\
&= \left[\sum_{m=1}^n \frac{\partial^2 f_k}{\partial x_m^2}; 1 \leq k \leq n \right] = \left[ \Delta f_k; 1 \leq k \leq n \right] = \Delta f(x). \label{eq_laplace_identity1}
\end{split}
\end{align}
The first two terms cancel out because of the symmetry of second derivatives (Schwarz's theorem) and interchange of sum and derivative.
\end{proof}

\begin{conj}[Operator Identity: $(-1)^n \curl \overline{\curl} f = \ROT \overline{\ROT} f$] \label{conj_laplace_identity2}
For any twice continuously differentiable vector field $f \in C^2(\mathbb{R}^n, \mathbb{R}^n)$, the following identity holds:
\begin{align}
(-1)^n \curl \overline{\curl} f(x) &= \ROT \overline{\ROT} f(x).
\end{align}
\end{conj}
\begin{proof}
This follows from Proposition~\ref{conj_laplace_identity1} and the well-known identity \citep[p.~522]{de_la_calle_ysern_constructive_2019}
\begin{align}
\Delta f(x) = \grad \Div f(x) + (-1)^n \curl \overline{\curl} f(x).
\end{align}
\end{proof}

\begin{conj}[Identity for Newton Potential $F$: $\overline{\mathcal{D}} \mathcal{D} F = \Delta F$] \label{eq_laplace_identity3} 
For any twice continuously differentiable vector field $f \in C^2(\mathbb{R}^n, \mathbb{R}^n)$ that decays faster than $\left|x\right|^{-c}$ for $\left|x\right| \to \infty$ and $c>0$, the potential $F(x) = \newtonintegralt \overline{\mathcal{D}} f(x)$ as defined by Eqs.~\eqref{eq_def_overlineD} and \eqref{eq_def_F} satisfies the following identity:
\begin{align}
\overline{\mathcal{D}} \mathcal{D} F(x) = \Delta F(x).
\end{align}
\end{conj}
\begin{proof} Using the definition of $F(x)$, Corollary \ref{sec_corol} stating $\newtonintegralt \overline{\mathcal{D}} f(x) = \overline{\mathcal{D}} \,\newtonintegralt f(x)$, Proposition \ref{conj_laplace_identity1} stating $\mathcal{D} \overline{\mathcal{D}} = \Delta$, the definition of $\phi(x)$ in Eq.~\eqref{eq_def_overlineD}, and $\Delta F(x) = \phi(x)$ from Eq.~\eqref{eq_laplace_identities_newton}:
\begin{align}
\overline{\mathcal{D}} \mathcal{D} F(x) = \overline{\mathcal{D}} \mathcal{D} \,\newtonintegral \overline{\mathcal{D}} f(x) = \overline{\mathcal{D}} \mathcal{D} \overline{\mathcal{D}} \,\newtonintegral f(x) = \overline{\mathcal{D}} \Delta \newtonintegral f(x) = \overline{\mathcal{D}} f(x) = \phi(x) = \Delta F(x).
\end{align}
\end{proof}
\noindent \textbf{Note:} While Proposition~\ref{conj_laplace_identity1} is an operator identity for any field $f(x)$, this derivation is valid only if $F(x)$ was constructed following Eqs.~\eqref{eq_def_overlineD} and \eqref{eq_def_F}. For a general function $Q \in C^3(\mathbb{R}^n, \mathbb{R}^{1+n^2})$, $\overline{\mathcal{D}} \mathcal{D} Q(x)$ may differ from $\Delta Q(x)$.

\begin{conj}[Eqs.~(\ref{eq_hdt3}--\ref{eq_hdt4}) of the Helmholtz Decomposition Theorem: $g(x) + r(x) = f(x)$] \label{conj_HT3} 

For any twice continuously differentiable vector field $f\colon \mathbb{R}^n \to \mathbb{R}^n$ that decays faster than $\left|x\right|^{-c}$ for $\left|x\right| \to \infty$ and $c>0$, the source potential $G(x) = \newtonintegralt \Div f(x)$ as defined by Eq.~\eqref{eq_newton_G} and its gradient $g(x) = \grad G(x)$, the rotation potential $R(x) = \newtonintegralt \overline{\ROT} f(x)$ as defined by Eqs.~\eqref{eq_newton_Rij} and \eqref{eq_newton_R} and its rotation $r(x) = \ROT R(x)$, and the potential $F(x) = \{ G(x), R(x) \}$ as defined by Eq.~\eqref{eq_def_F} satisfy the following identity:
\begin{align}
g(x) + r(x) = \grad G(x) + \ROT R(x) = \mathcal{D} F(x) = f(x). \label{eq_HI1}
\end{align}
\end{conj}
\begin{proof} The first equalities follow immediately from the definitions. The last can be derived starting with the definition of $F(x)$, applying Corollary~\ref{sec_corol} and Proposition \ref{conj_laplace_identity1}:
\begin{align}
\mathcal{D} F(x) = \mathcal{D} \,\newtonintegral \overline{\mathcal{D}} f(x) = \mathcal{D} \overline{\mathcal{D}} \,\newtonintegral f(x) = \Delta \newtonintegral f(x) = f(x).
\end{align}
\end{proof}

This completes the proof of the Helmholtz Decomposition Theorem.

\section{Conclusions}

\label{sec_conclusions}

In this paper, we have introduced differential operators $\overline{\ROT}$, $\ROT$ and the generalized derivatives $\overline{\mathcal{D}}$ and $\mathcal{D}$, such that for any twice continuously differentiable vector field $f(x)$ in $\mathbb{R}^n$ that decays faster then $\left|x\right|^{-c}$ for $x \to \infty$ and $c>0$, a scalar source potential $G(x)$ and an antisymmetric rotation potential $R(x) = R_{ij}(x)$ with $\binom{n}{2}$ distinct entries can be calculated as convolutions of the density $\phi(x) = \overline{\mathcal{D}} f(x)$ with the fundamental solutions of the Laplace equation.
The joint potential $F(x) = \left\{ G(x), R(x) \right\}$ is an antiderivative of $f(x)$, such that applying the differential operator $\mathcal{D}$ to this potential provides a decomposition of $f(x)$ into a rotation-free `gradient field' $g(x) = \grad G(x)$ and a source-free `rotation field' $r(x) = \ROT R(x)$:
\begin{align}
f(x) & = \mathcal{D} F(x) = \mathcal{D} \,\newtonintegralt \overline{\mathcal{D}} f(x) 
 = \mathcal{D} \,\newtonintegral \phi(x) = g(x) + r(x).
\end{align}
This generalizes the Helmholtz Decomposition to $\mathbb{R}^n$ without the need for differential forms and the complicated operator $\curl$ and facilitates its application to high-dimensional dynamic systems.
The potentials correspond to `antiderivatives' of the gradient and rotation differential operators, providing a generalization of the fundamental theorem of calculus in $\mathbb{R}^n$ that links differential and integral calculus.

\clearpage

\section*{Acknowledgments}
EG thanks Ulf Klein and Walter Zulehner from Johannes Kepler University Linz. OR thanks Ulrike Feudel and Jan Freund from Carl von Ossietzky University of Oldenburg, and Anja Janischewski.

\renewcommand{\bibfont}{\normalfont\small}
\addcontentsline{toc}{section}{References} 

\printbibliography

\newgeometry{inner=1.cm,outer=1.cm,top=0.9cm,bottom=0.cm,includeheadfoot , heightrounded}

\enlargethispage{2cm}

\begin{appendices}

\section{Comparison of the Helmholtz Decomposition, operators and variables in dimension n with dimension 3}

\label{tabelle-dim-n-3}

{

\footnotesize

\setlength{\abovedisplayskip}{-1em}
\setlength{\belowdisplayskip}{-0.5em}

\begin{tabular}{L{0.17\columnwidth} L{0.33\columnwidth} L{0.4\columnwidth}}
\toprule
     \textbf{Variable or operator}  & \textbf{dimension n}   &  \textbf{dimension 3} 
\\ \midrule  
     source potential $G \in C^3(\mathbb{R}^n, \mathbb{R})$
  &  \begin{align*} G(x) \end{align*}
  &  \begin{align} \Phi(x) = -G(x) \end{align}
\\ \midrule
     basic rotation potentials $R_{ij}, R_i \in C^3(\mathbb{R}^n, \mathbb{R})$
  &  \begin{align*} R_{ij}(x) = -R_{ji}(x); 1\leq i, j \leq n, \qquad \tbinom{n}{2} \text{ distinct entries } \end{align*}
  &  \begin{align} R_1, R_2, R_3, \qquad \tbinom{n}{2} = 3 \text{ distinct entries}  \end{align}
\\ \midrule
     rotation potential \newline $R \in C^3(\mathbb{R}^n, \mathbb{R}^{n^2})$
  &  \begin{align*} R(x) = \left\llbracket R_{ij}(x); 1\leq i, j \leq n \right\rrbracket  \qquad n \times n \text{ matrix} \end{align*} 
  &  \begin{align} R(x) = A(x) = \left[ R_1, R_2, R_3 \right] \end{align}
\\ \midrule
     potential $F \in C^3(\mathbb{R}^{n}, \mathbb{R}^{1+n^2})$ 
  &  \begin{align*} F(x) = \Big\{G(x), R(x)\Big\} \end{align*}
  &  \begin{align} F(x) = \Big\{G(x), R(x)\Big\}  \end{align}
\\ \midrule
     basic rotation operator $\ROT_{ij}\colon C^1(\mathbb{R}^n,\mathbb{R}) \to C^0(\mathbb{R}^n,\mathbb{R}^n)$ 
  &  \begin{align*} \ROT_{ij} R_{ij} \coloneqq \left[ \delta_{ik} \frac{\partial R_{ij}}{\partial x_j} - \delta_{jk} \frac{\partial R_{ij}}{\partial x_i}; 1 \leq k \leq n \right] \end{align*}
  &  \begin{align}  \end{align}
\\ \midrule
     rotation operator $\ROT R\colon C^1(\mathbb{R}^n,\mathbb{R}^{n^2}) \to C^0(\mathbb{R}^n,\mathbb{R}^n)$   
  &  { \begin{align*} \ROT R \coloneqq & \sum_{1\leq i, j \leq n} \tfrac12 \ROT_{ij} R_{ij} = \left[ \sum_{m} \frac{\partial R_{km}}{\partial x_m}; 1 \leq k \leq n \right] \end{align*} }
  &  \begin{align} \curl R = \left[ \frac{\partial R_3}{\partial x_2} - \frac{\partial R_2}{\partial x_3}, \ \frac{\partial R_1}{\partial x_3} - \frac{\partial R_3}{\partial x_1}, \ \frac{\partial R_2}{\partial x_1} - \frac{\partial R_1}{\partial x_2} \right] \end{align}   \\
  &  \multicolumn{2}{r}{ $\ROT R = \curl R$ with $R_1 = R_{23}$; $R_2 = - R_{13}$; $R_3 = + R_{1,2} \quad$ }
\\ \midrule
     vector field $f$, gradient field $g$ and rotation field $r$  \newline $f, g, r \in C^2(\mathbb{R}^n, \mathbb{R}^n)$
  &  {\begin{align*} f(x) &= g(x) + r(x) \\ g(x) &= \grad G(x); \quad r(x) = \ROT R(x) \end{align*}}
  &  {\begin{align} f(x) &= g(x) + r(x)  \\ g(x) &= \grad G(x); \quad r(x) = \curl R(x) \end{align}}
\\ \midrule
     gradient fields are rotation-free
  &  \begin{align*} \overline{\ROT} g(x) = \overline{\ROT} \grad G(x) = 0 \end{align*}
  &  \begin{align} \overline{\curl} g(x) = \overline{\curl} \grad G(x) = 0 \label{tab_eq_grad_no_rot} \end{align}
\\ \midrule
     rotational fields are divergence-free
  &  \begin{align*} \Div r(x) = \Div \ROT R(x) = 0 \end{align*}
  &  \begin{align} \Div r(x) = \Div \curl R(x) = 0 \label{tab_eq_rot_no_div} \end{align}
\\ \midrule
     derivative of a potential $\mathcal{D}\colon C^1(\mathbb{R}^n,\mathbb{R}^{1+n^2}) \to C^0(\mathbb{R}^n,\mathbb{R}^n)$  
  &  \begin{align*} f(x) = \mathcal{D} F(x) \coloneqq \grad G(x) + \ROT R(x) \end{align*}
  &  \begin{align} f(x) = \mathcal{D} F(x) \coloneqq - \grad \Phi(x) + \curl R(x) \end{align}
\\ \midrule
     scalar source density $\gamma$ \newline $\gamma \in C^1(\mathbb{R}^n, \mathbb{R})$ 
  &  \begin{align*} \gamma(x) = \Div f(x) = \Delta G(x) \end{align*}
  &  \begin{align} \gamma(x) = \Div f(x) = \Delta G(x) \end{align}
\\ \midrule
     basic rotation density op. \newline $\overline{\ROT_{ij}}\colon C^1(\mathbb{R}^n,\mathbb{R}^n) \to C^0(\mathbb{R}^n,\mathbb{R})$ 
  &  \begin{align*} \overline{\ROT_{ij}} f(x) \coloneqq \left( \frac{\partial f_i}{\partial x_j} - \frac{\partial f_j}{\partial x_i} \right) \end{align*}
  &  \begin{align} \end{align}
\\ \midrule
     rotation density operator $\overline{\ROT}\colon C^1(\mathbb{R}^n,\mathbb{R}^n) \to C^0(\mathbb{R}^n,\mathbb{R}^{n^2})$   
  &  \begin{align*} \overline{\ROT} f(x) \coloneqq \left\llbracket \overline{\ROT_{ij}} f(x) \right\rrbracket = \left\llbracket \left( \frac{\partial f_i}{\partial x_j} - \frac{\partial f_j}{\partial x_i} \right); 1 \leq i, j \leq n \right\rrbracket \end{align*}
  &  \begin{align} \overline{\curl} f = \left[ \frac{\partial f_3}{\partial x_2} - \frac{\partial f_2}{\partial x_3}, \ \frac{\partial f_1}{\partial x_3} - \frac{\partial f_3}{\partial x_1}, \ \frac{\partial f_2}{\partial x_1} - \frac{\partial f_1}{\partial x_2} \right] \end{align}
\\ \midrule
     basic rotation density $\rho_{ij}, \rho_i \in C^1(\mathbb{R}^n, \mathbb{R})$
  &  \begin{align*} \rho_{ij}(x) = - \rho_{ji}(x) = \overline{\ROT_{ij}} f(x); 1 \leq i, j \leq n, \\ \tbinom{n}{2} \text{ distinct entries} \nonumber \end{align*}
  &  \begin{align} \rho_1(x), \rho_2(x), \rho_3(x) \\ \tbinom{n}{2} = 3 \text{ distinct entries}   \end{align}
\\ \midrule
     rotation density $\rho$ \newline $\rho \in C^1(\mathbb{R}^n, \mathbb{R}^{n^2})$
  &  \begin{align*} \rho(x) = \left\llbracket \rho_{ij}(x); 1 \leq i, j \leq n \right\rrbracket = \overline{\ROT} f(x) = \overline{\ROT} \ROT R \end{align*} \newline $n \times n$ matrix
  &  \begin{align}\rho(x) = \left[ \rho_1, \rho_2, \rho_3 \right] = \overline{\curl} f(x)  = \overline{\curl} \curl R(x) \end{align} \newline \hspace*{1.5em} identify: $\rho_1 = - \rho^{2,3}$; $\rho_2 = + \rho^{1,3}$; $\rho_3 = - \rho^{1,2}$
\\ \midrule
     density derivative $\overline{\mathcal{D}}\colon C^1(\mathbb{R}^n,\mathbb{R}^n) \to C^0(\mathbb{R}^n,\mathbb{R}^{1+n^2})$
  &  \begin{align*} \overline{\mathcal{D}} f(x) \coloneqq \left\{ \Div f(x), \overline{\ROT} f(x) \right\} \end{align*}
  &  \begin{align} \overline{\mathcal{D}} f(x) \coloneqq \left\{ \Div f(x), \overline{\curl} f(x) \right\} \end{align}
\\ \midrule
     density \newline $\phi \in C^1(\mathbb{R}^n, \mathbb{R}^{1+n^2})$
  &  \begin{align*} \phi(x) = \big\{\gamma(x), \rho(x)\big\} = \left\{\gamma(x), \left\llbracket \rho_{ij}(x) \right\rrbracket \right\} = \overline{\mathcal{D}} f(x) \end{align*}
  &  \begin{align} \phi(x) = \big\{\gamma(x), \rho(x)\big\} =  \left\{\gamma(x), \left[ \rho_1(x), \rho_2(x), \rho_3(x) \right] \right\} = \overline{\mathcal{D}} f(x) \end{align}
\\ \midrule
Newton potential operator \newline $\newtonintegralt\colon C^0(\mathbb{R}^n,\mathbb{R}) \to C^2(\mathbb{R}^n,\mathbb{R})$ %
  &  { \begin{align*} G(x) &= \newtonintegral \gamma(x) = \int_{\mathbb{R}^n} K(x, \xi)\ \gamma(\xi)\ d\xi^n \\ R_{ij}(x) &= \newtonintegral \rho_{ij}(x) = \int_{\mathbb{R}^n} K(x, \xi)\ \rho_{ij}(\xi)\ d\xi^n \nonumber \end{align*} }
  &  { \begin{align} \Phi(x) &= \frac{1}{4\pi} \iiint_{\mathbb{R}^3}  \frac{\gamma(\xi) }{|x-\xi|} \,d\xi^3 \\ R(x) &= \frac{1}{4\pi}\iiint_{\mathbb{R}^3} \frac{\rho(\xi) }{|x-\xi|} \,d\xi^3 \end{align} }
\\[0.8em]
     \multicolumn{3}{l}{with the kernel $K(x,\xi) = \begin{cases} 
         \frac{1}{2\pi} (\log{ | x-\xi | } - \log{ | \xi | }) &  n=2  \\
         \frac{1}{n(2-n)V_n} (| x-\xi | ^{2-n} - | \xi | ^{2-n}) &  n \neq 2,\text{ and } V_n = \pi^\frac{n}{2} / \Gamma\big(\tfrac{n}{2}+1\big) \text{ the volume of a unit $n$-ball, } V_3 = 4\pi/3
      \end{cases}$}         
\\ \bottomrule
\end{tabular}
}


\end{appendices}

\end{document}